\documentclass[10pt,letterpaper]{article}
\usepackage{opex3}

\usepackage{amsmath}
\usepackage[tight]{units}

\makeatletter
\DeclareRobustCommand*\textsubscript[1]{%
\@textsubscript{\selectfont#1}}
\newcommand{\@textsubscript}[1]{%
{\m@th\ensuremath{_{\mbox{\fontsize\sf@size\z@#1}}}}}
\makeatother

\newcommand{\mysection}[1]{}

\newlength{\myfigwidth}
\newlength{\myfigwidthtwo}
\newcommand{\tsubs}[1]{\textsubscript{#1}}

\newcommand{\mytextfrac}[2]{#1/#2}
\newcommand{\rmd}{\mathrm{d}}

\begin{document}

\title{Ultrashort pulse characterization by spectral shearing interferometry with spatially chirped ancillae}




\author{Tobias Witting,$^{\dagger}$ Dane R. Austin$^\dagger$, and Ian
  A. Walmsley}
\address{Clarendon Laboratory, University of Oxford,
  Parks Road, Oxford, OX1 3PU, UK}
\email{t.witting@physics.ox.ac.uk}

\begin{abstract}
  We report a new version of spectral phase interferometry for direct electric field reconstruction (SPIDER), which
  enables consistency checking through the simultaneous acquisition of multiple shears and offers a simple and precise
  calibration method. By mixing the test pulse with two spatially chirped ancilla fields we generate a single-shot
  interferogram which contains multiple shears, the spectral amplitude of the test pulse, and the reference phase, which
  is accurate for broadband pulses. All calibration parameters --- shear, upconversion-frequency and reference phase
  position --- can be accurately obtained from a single calibration trace.
\end{abstract}


\bibliographystyle{osajnl}

\section{Introduction}
\label{sec:introduction}
Ultrashort pulse characterization~\cite{Walmsley_2009_AOP_Characterization-of-} is a key technology underpinning much of
ultrafast science. Progress in the field continues to occur on many fronts --- extending the range of measurable
pulse parameters, such as duration, wavelength, and complexity, improving the sensitivity and single-shot capability,
and providing spatial, in addition to temporal, resolution.  Less easily quantified but equally important issues are
precision, robustness, ease of use and the diagnosis, identification and quantification of errors. Innovations on this
front have included simplified geometrical
setups~\cite{OShea_2001_OL_Highly-simplified-de,Radunsky_2007_L_Compact-spectral-shearing-inte}, theoretical analysis of
the influence of various errors~\cite{Dorrer_1999_JOSAB_Influence-of-the-calibration-o}, and the exploitation of
redundant data to identify misalignment or violations of the assumptions on the pulse
properties~\cite{Dorrer_2002_JOSAB_Precision-and-consistency-crit,Trebino_2002__Frequency-resolved-o}.  Spectral
shearing interferometry (SSI), a popular class of techniques whose best-known example is Spectral Phase Interferometry
for Direct Electric field Reconstruction (SPIDER)~\cite{Iaconis_1998_OL_Spectral-phase-interferometry-} is widely used
for its efficient and direct retrieval acquisition and computation requirements, which scale linearly with the
time-bandwidth product of the unknown pulse, and its natural suitability to broadband optical input. Its direct
retrieval formula makes analysis of the influence of various errors a straightforward matter and its one dimensional
data encoding makes the acquisition of multiple shots, and subsequent analysis of their shot-to-shot variations,
convenient and rapid. Geometrically simplified arrangements have also been
developed~\cite{Radunsky_2007_L_Compact-spectral-shearing-inte}. The common feature of all SSI techniques is the
interference of two spectrally sheared replicas of the unknown pulse. This introduces two parameters, the shear and the
interferometric reference phase, both of which must be calibrated to a precision which increases with the time-bandwidth
product of the unknown pulse. Novel geometries for the convenient determination of these
parameters~\cite{Kosik_2005_OL_Interferometric-technique-for-,%
  Birge_2006_OL_Two-dimensional-spectral-shear,%
  Witting_2009_OL_Improved-ancilla-pre} are therefore of interest, supplementing the well-established
techniques~\cite{Iaconis_1998_OL_Spectral-phase-interferometry-}.  Additionally, whilst typical raw SSI data contains
some redundant information~\cite{Dorrer_2002_JOSAB_Precision-and-consistency-crit} enabling self-consistency checking,
this can be supplemented by taking several measurements at different shears which should result in identical retrieved
fields.

Most currently known SPIDER variants~\cite{%
  Iaconis_1998_OL_Spectral-phase-interferometry-,%
  Kosik_2005_OL_Interferometric-technique-for-,%
  Birge_2006_OL_Two-dimensional-spectral-shear,%
  Monmayrant_2003_OL_Time-domain-interferometry-for,%
  Baum_2004_OL_Zero-additional-phase-SPIDER:-,%
  Lelek_2006_OC_Time-resolved-spectr} are implemented as single shear devices. Changing the shear involves moving parts
which requires additional calibration and prevents single shot multiple shear operation.  A significant recent
development in this context  is Chirped Ancillae aRrangement- (CAR-)SPIDER~\cite{Gorza_2007_OE_Spectral-shearing-in}, in
which an upconverted replica of the unknown pulse with spatially varying frequency is produced through the phasematching properties
of a long crystal. This replica is split, and two-dimensional spectral interferometry is performed between itself and a
laterally inverted copy, yielding an interferogram in which the shear varies with position. The zero-shear line encodes
the reference phase, whilst the shear calibration can be performed by blocking each arm individually. CAR-SPIDER is
therefore an attractive device in the light of the previous discussion.  However, the collinear long-crystal geometry limits
both the measureable  bandwidth and the range of pulse lengths. Also in the long-crystal arrangement
the spatial carrier of the interferogram depends on frequency, requiring an additional calibration for broadband pulses.

In this paper, we present a device which we dub Spatially Encoded Arrangement for CAR-SPIDER (SEA-CAR-SPIDER), in which
spatially chirped ancillae are produced by a diffraction grating and split before undergoing upconversion with the
test pulse. This removes the restriction on the pulse parameters imposed by the long-crystal geometry and also renders
the spatial carrier independent of frequency, so that the reference phase is truly self-calibrated.  Furthermore, the
arrangement permits a convenient determination of all calibration parameters in a single step with much greater
precision than CAR-SPIDER, as well as a direct verification of the ancillae monochromaticity which is important for
accurate measurement of long pulses.

\section{SEA-CAR-SPIDER theory}
\label{sec:sea-car-spider}

The test pulse $E(\omega)=\sqrt{I(\omega)}|\exp[i\phi(\omega)]$, propagating along the $z$-axis, undergoes sum-frequency
generation with two spatially chirped ancillae, one of which has been laterally inverted about the $y$-axis, so that
their local frequency is $\omega =\omega\tsubs{up}\pm \alpha x$, where $\omega\tsubs{up}$ is the upconversion frequency
common to both ancillae at $x=0$. The ancillae wavevectors also form an angle $\vartheta$ in the $x$-$z$-plane,
symmetrically about the $z$-axis. The resulting interferogram, re-imaged onto two-dimensional spectrometer, is of the
form
\begin{equation}
  S(\omega+\omega\tsubs{up},x) = I(\omega+\alpha x) + I(\omega-\alpha x) +
  D(\omega,x) + D^\ast(\omega,x)
\end{equation}
\begin{equation}
  D(\omega,x) = \sqrt{I(\omega+\alpha x])I(\omega-\alpha x)}\exp \left[ i \Gamma(\omega,x) \right]
\end{equation}
\begin{equation}
  \Gamma(\omega,x) = \phi(\omega+\alpha x) -
  \phi(\omega-\alpha x)+
  \label{eq:gamma}
  k_x x + C(\omega)
\end{equation}
where $D(\omega,x)$ is the positive-spatial-frequency AC sideband with spatial carrier $k_{x} =
\mytextfrac{\omega\tsubs{up}\vartheta}{c}$, $\Gamma(\omega,x)$ is the interferogram phase, and $C(\omega)$ is the
zero-shear calibration phase, which is theoretically zero due to symmetry but may be present due to imperfect imaging.
Note that the spatial carrier $k_x$ is independent of the signal frequency $\omega$ because the spatial tilt is applied
to the ancillae, rather than the upconverted replica as in CAR-SPIDER.  At all $x$-positions $x\neq0$, the interferogram phase
contains the finite differences $\phi(\omega+\alpha x) - \phi(\omega-\alpha x)$ i.e. spectral shearing interferometry data of shear $\Omega=2\alpha x$.  

The reconstruction procedure is as follows:  the
phase-gradient $\Gamma(\omega,x)$, contained in the AC sideband at spatial frequency $k_{x} =
\mytextfrac{\omega\tsubs{up}\vartheta}{c}$, is extracted by 2D Fourier filtering, as with
SEA-SPIDER~\cite{Kosik_2005_OL_Interferometric-technique-for-}.  Then the calibration phase must be determined.  If the
spectrum has no nulls, then $\Gamma(\omega,x)$ is well defined along $x = 0$ and $C(\omega)$ can be obtained directly at
this position and subtracted from $\Gamma(\omega,x)$~\cite{Gorza_2007_OE_Spectral-shearing-in}.  Alternatively, in the
case of spectral nulls at $x=0$, or for increased precision, the calibration phase can also be obtained by weighted averaging
along the $x$ axis. We use the fact that, except for the calibration phase, $\Gamma(\omega,x)$ is an odd function about $x=0$.
Therefore by averaging accross a symmetric interval about $x=0$ the calibration phase can be isolated. In practice this
requires a method, which is robust against $2\pi$ phase discontinuities. We use
\begin{equation}
  \label{eq:}
  C(\omega) = \frac{1}{2} \text{unwrap}\left[\text{Arg} \int_0^{\infty} D(\omega,x) D(\omega,-x) \rmd x \right]
\end{equation}
where $\text{Arg}$ is the principal value of the argument and $\text{unwrap}$ denotes a one-dimensional phase unwrapping routine which removes $2\pi$ discontinuities. 

After subtracting the calibration phase, the standard concatenation procedure can be performed at any $x$ as
desired. The spectral amplitude of the unknown pulse, modulated by the phasematching efficiency and the spectrometer
response, is also obtained from the $x=0$ signal. Therefore, all data is recorded near the upconversion frequency,
enabling a specificially optimized spectrometer to be used. Even though a single SEA-CAR-Spidergram contains tens of
independent datasets, online reconstruction of each of these is still possible owing to the fast algebraic reconstruction
algorithm of SPIDER.

We note that as with filter-SPIDER~\cite{Witting_2009_OL_Improved-ancilla-pre} and CAR-SPIDER, our new device does not
rely on temporal stretching of the ancillae to ensure the quasi-monochromaticity.  Thus in SEA-CAR-SPIDER, which is free from
the pulse length restrictions of the long-crystal arrangement, strongly chirped pulses can be characterized without the
need for correction of the reconstructed phase~\cite{Wemans_2006_OL_Self-referencing-spe}.

\begin{figure}[htb]
  \centering
  \includegraphics[width=0.9\textwidth]{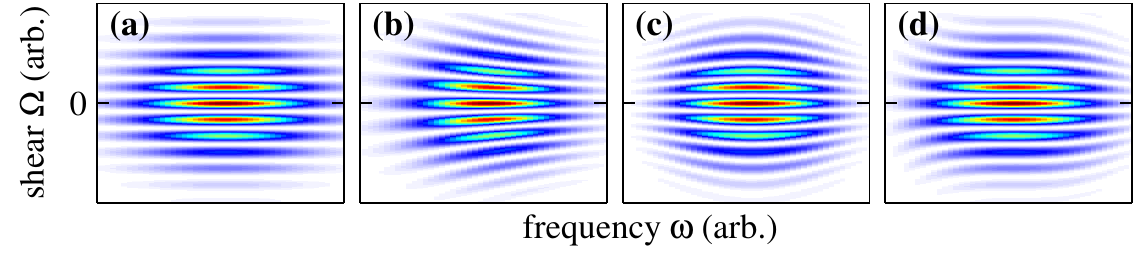}
  \caption{Example SEA-CAR-SPIDER traces calculated for pulses with different order polynomial spectral phase. (a)
    transform-limited pulse; (b) quadratic spectral phase; (c) cubic spectral phase; (d) quartic spectral phase. The
    fringes map out the phase gradient scaled by the shear.}
 \label{fig:seacartracestheory}
\end{figure}

The spatial carrier in SEA-CAR-SPIDER enables an intuitive interpretation of the fringe pattern, because, as we now
show, the fringe contours follow the phase derivative of the unknown pulse.  Let $x\tsubs{con}(\omega)$ be a fringe
contour, defined by $\Gamma=\text{constant}$. Then
\begin{equation}
  \label{eq:fringecontourdef}
  \frac{\rmd \Gamma[\omega, x\tsubs{con}(\omega)]}{\rmd \omega}=
  \frac{\partial \Gamma[\omega, x\tsubs{con}(\omega)]}{\partial \omega} +
  \frac{\partial \Gamma[\omega,x\tsubs{con}(\omega)]}{\partial x}  \frac{\partial x\tsubs{con}(\omega)}{\partial \omega}=0.
\end{equation}
In~(\ref{eq:gamma}) we approximate the finite differences with a derivative, and write
\begin{equation}
  \label{eq:derivative}
  \Gamma(\omega,x)\approx\frac{\partial \phi(\omega)}{\partial \omega} 2 \alpha x + k_x x
\end{equation}
The spatial derivative is
\begin{equation}
  \label{eq:derivative2}
  \frac{\partial \Gamma(\omega,x)}{\partial x} = \frac{\partial \phi(\omega)}{\partial \omega} 2
  \alpha + k_x .
\end{equation}
The second term on the right hand side of \eqref{eq:derivative2} represents the spatial carrier and is much larger than
the first term --- otherwise the sidebands would not be distinct in the Fourier domain. We neglect the first term and
substitute (\ref{eq:derivative2}) into (\ref{eq:fringecontourdef}), yielding
\begin{equation}
  \label{eq:2}
  \frac{\partial x\tsubs{con}(\omega)}{\partial \omega} =
  -\frac{2\alpha x\tsubs{con}(\omega)}{k_x}
  \frac{\partial^2\phi}{\partial \omega^2}.
\end{equation}
Therefore the fringe contours follow the spectral phase gradient scaled by the local shear and the inverse of the
spatial carrier. This is illustrated in Fig.~\ref{fig:seacartracestheory}, which shows calculated traces for a
transform-limited pulse and for pulses with quadratic, cubic and quartic polynomial spectral phases. Note that for a transform-limited pulse (Fig.~\ref{fig:seacartracestheory}~(a)) the fringe contours are flat, reflecting the frequency
independence of the spatial carrier. The fringe pattern provides a simple and useful visual tool for aligning
a laser qualitatively without the need for running a reconstruction algorithm.

\section{Experimental Setup}
\label{sec:experimental-setup}
\begin{figure}[htb]
  \centering
  \includegraphics[width=0.7\textwidth]{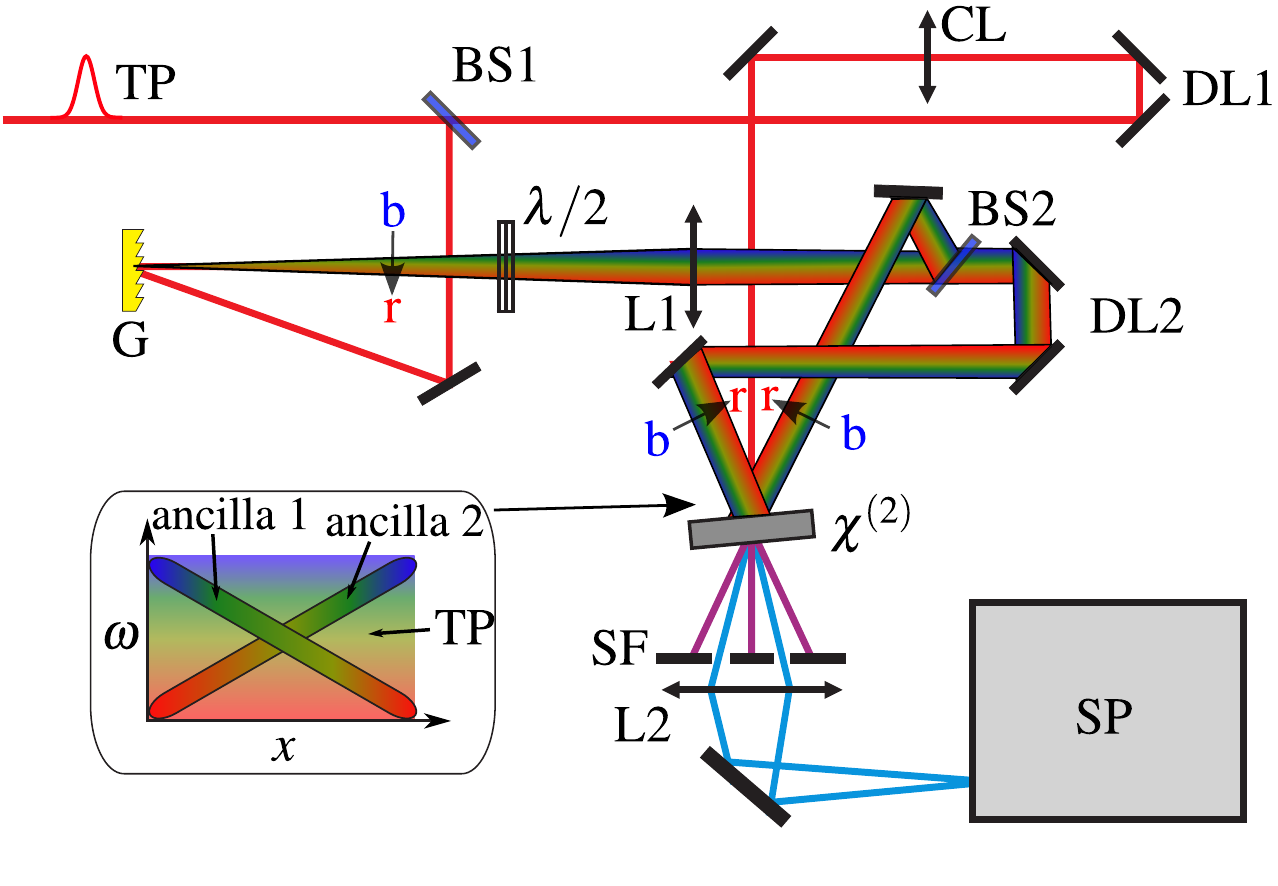}
  \caption{SEA-CAR-SPIDER experimental setup. For detailed
    explanation, refer to text.}
  \label{fig:schematicsetup}
\end{figure}

The experimental implementation of SEA-CAR-SPIDER is depicted in Fig.~\ref{fig:schematicsetup}. The test pulse TP is
split with beamsplitter BS1 and focussed into the nonlinear crystal $\chi^{(2)}$ with a cylindrical lens CL with focal
length $f\tsubs{cl}=\unit[500]{mm}$. The delay-line DL1 is only for coarse adjustment of the delay between ancillae and
test pulse and only requires initial adjustment.  The ancillae are prepared as follows: the reflected beam from BS1 is
dispersed by a grating of pitch $\Lambda = \unit[300]{mm^{-1}}$ and angle of incidence $\gamma=\unit[10]{^{\circ}}$. The
spherical lens L1 is positioned one focal length $f_1 = \unit[500]{mm}$ away from the grating and the crystal is mounted
$f_1$ further downstream. Between the lens and the crystal the ancilla beam is split at 50/50 beamsplitter BS2 and one
arm is spatially inverted. The coarse delay line DL2 is used to achieve temporal overlap between the ancillae and needs
no subsequent adjustment.  The $\chi^{(2)}$ crystal (\unit[250]{\textmu m} BBO) is placed in the back fourier-plane of
L1, where the spatial chirp of the ancillae is $\alpha = \cos{[\beta(\omega\tsubs{up})]} \omega\tsubs{up}^2 / (2\pi c
f\tsubs{1} \Lambda )$ where $\beta(\omega\tsubs{up})$ is the diffracted angle of the beam from the grating. A half-wave
retarder $\lambda/2$ is introduced into the ancilla arm to ensure type-II upconversion with the ancilla polarization
aligned to the $e$-axis and the TP polarization to the $o$-axis of the crystal, respectively.  After spatial filtering
(SF) the sum-frequency beams are re-imaged onto a 2D imaging spectrometer by lens L2 with focal length
$f_2=\unit[300]{mm}$.  The spectrometer's entrance slit is oriented in the dispersion plane of the ancillae and the
signal is internally dispersed perpendicular to this plane.  The spectrometer provides astigmatism-free imaging over
wide spectral and spatial extent~\cite{Austin_2009_AO_Broadband-astigmatisA}. The signal is recorded with a
$\unit[1280\times1024]{pixel}$ CMOS detector with \unit[8]{bit} ADC resolution.  Single shot data is recorded at a rate
of \unit[30]{Hz} limited by memory and data-transfer capacities.

It should be noted that SEA-CAR-SPIDER can be operated as zero
additional phase (ZAP) device by using a cylindrical mirror and reflection of the test-pulse off BS1 rather than
transmission.

\section{Calibration}
\label{sec:calibration}
In order to obtain a single-step calibration of the shear slope $\alpha$, as well as the central upconversion frequency
$\omega\tsubs{up}$ and the $x$-origin, the TP arm was blocked and the spatial filter SF temporarily removed. In this
configuration, the second harmonic of the ancillae, as well as their sum-frequency signal, are obtained.
\begin{figure}[htb]
\centering
 \includegraphics{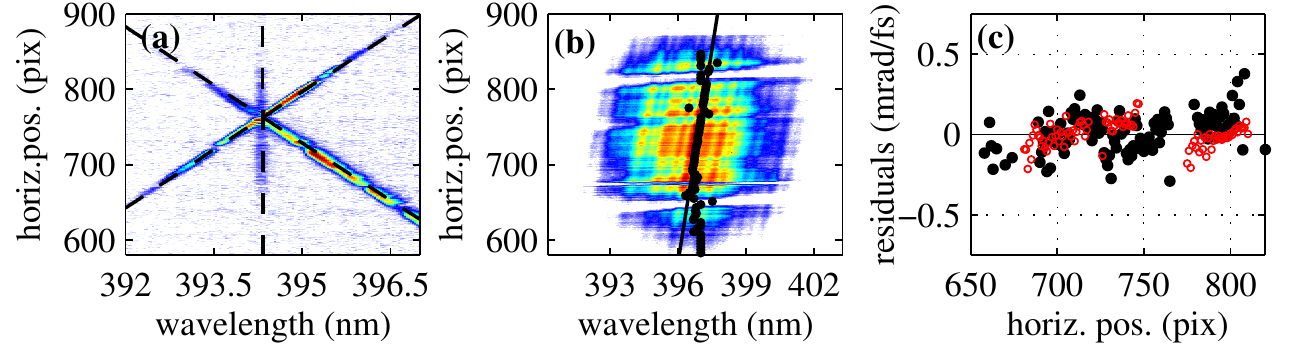}
  \caption{SEA-CAR-SPIDER calibration: (a) Spatially resolved spectrum of the upconversion from the two ancilla pulses
    yields calibration of upconversion frequency and shear. 20\,dB color scale. The black dashed lines are linear
    fits. (b) fitting the shear using the sum-frequency signal. Central freqencies of individual spectral slices (black
    dots) obtained from cross-correlation, linear fit (black line). (c) residuals of fits of the ancilla as in (a) in
    red circles, residuals of fit (b) in black dots.}
 \label{fig:calibration}
\end{figure}
Fig.~\ref{fig:calibration}\,(a) shows the calibration data obtained from the average of 50 single shot acquisitions. A
linear fit of the slopes of the two ancillae provides $\alpha=\unit[22.93\pm0.06]{mrad/fs/mm}$.
From the crossing point of the two ancillae the $x$-origin and $\omega\tsubs{up}$ are obtained. 
We determine the $x$-origin to be pixel $762.3\pm0.1$. The upconversion frequency is found to be
$\omega\tsubs{up}=\unit[2.38767\pm2.39\times10^{-5}]{rad/fs}$. All errors are estimated using the covariances of the
individual fit parameters for both ancillae.

The sum-frequency mixing between the ancillae appears as the vertical line at frequency $\omega\tsubs{up}$ in
Fig.~\ref{fig:calibration}\,(a). Its position with respect to the crossing point is a useful check of the axial
alignment of the crystal-to-spectrometer reimaging system. If L2 is correctly relay imaging the crystal plane to the
spectrometer entrance slit, then all three lines cross at the same point.

The spectral width of the ancillae can also be obtained from the calibration data of
Fig.~\ref{fig:calibration}\,(a). This is useful in verifying that the ancillae are sufficiently monochromatic, a basic
requirement of spectral shearing interferometry. In the case of SEA-CAR-SPIDER, where many shears are available, the
spectral width of the ancillae determines the minimum shear from which a meaningful reconstruction can be
performed. Here, we obtain a monochromaticity of $\unit[1.85]{mrad/fs}$ FWHM (in the second harmonic or
$\unit[1.31]{mrad/fs}$ (in the fundamental), granting a maximal time window for test pulses of $T=\unit[4.8]{ps}$. The
ancillae spectral width is $\alpha s$, where $s$ is the FWHM spot size produced by lens L1. The monochromaticity can
therefore be further increased by using a higher pitched grating and a tighter focus.

We note that a similar calibration procedure was performed in the original CAR-SPIDER arrangement. However, because the
upconversion was performed using the asymmetric phasematching in a long crystal it is impossible to block the test pulse
and hence the shear calibration was performed using the upconverted signal pulses. As we now illustrate, the
SEA-CAR-SPIDER procedure offers greater precision because the second harmonics of the ancillae are much more narrowband
than the test pulse. In CAR-SPIDER, the calibration would be obtained by recording the individual signals separately and
obtaining their displacement by cross-correlation,  disregarding regions of low intensity.
One such fit is shown in Fig.~\ref{fig:calibration}\,(b). The residuals of the slope fits are directly compared in
Fig.~\ref{fig:calibration}~(c). Using this method we get shear slopes of
$\alpha\tsubs{sig}=\unit[22.76\pm2.5]{mrad/fs/mm}$. 
From the signal fits we further determine the $x$-origin to be $\unit[760.2\pm2.2]{pixels}$.  (The upconversion
frequency is not determined by this method).  As expected due to the greater spectral width of the signals the errors in
determining the calibration parameters are significantly larger compared to the new calibration procedure presented
above. 
This emphasizes the precision advantage of the calibration offered by the ancilla preparation in SEA-CAR-SPIDER.

\begin{figure}[htb]
  \centering
  \includegraphics[scale=1]{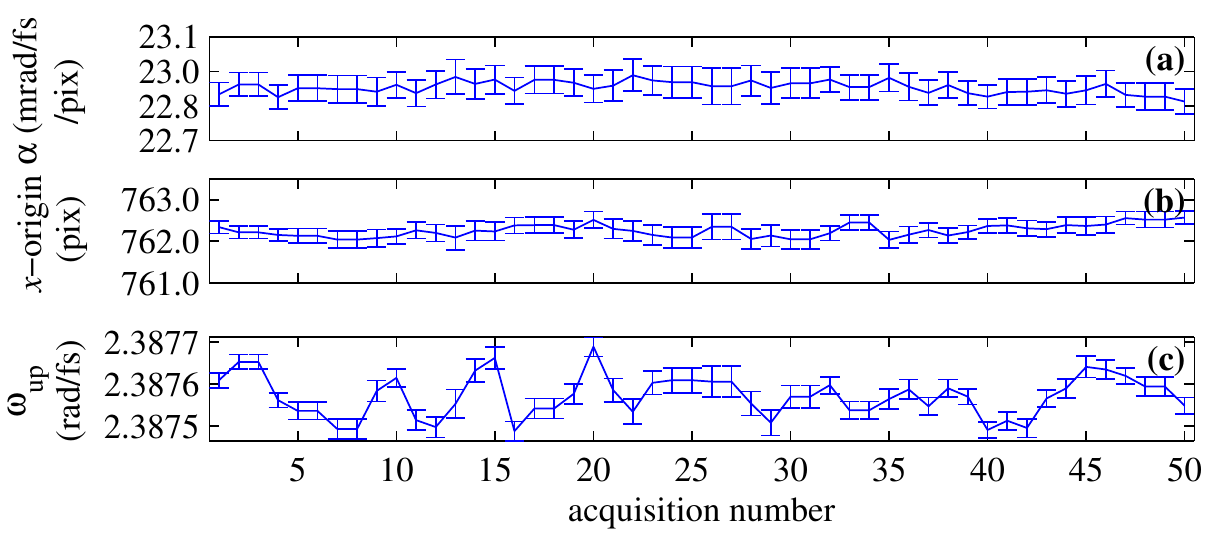}
  \caption{Evaluation of the stability of the calibration parameters. We recorded 50 single shot calibration traces and
    performed the fitting routines on each. (a) shear slope $\alpha$. (b) position of $x$-origin (zero shear row)
    determined from the crossing point. (c) upconversion frequency $\omega\tsubs{up}$.}
  \label{fig:calmultishots}
\end{figure}

We investigated the sensitivity of the calibration against fluctuations of the beam-pointing and displacement.  Our
device preserves vertical symmetry ($y$-axis in our coordinate system) and relay images all beams. It turns out that
beam-pointing and displacement changes in the vertical direction do not affect the device calibration. Vertical beam
pointing fluctuations cause the signal to move off the slit reducing the signal intensity.  Horizontal beam
displacements cause a change in the spatial carrier, which does not affect the calibration or reconstruction.
A beam-pointing change $\theta_x$ in the horizontal plane causes the two spatially chirped ancilla fields to move into
opposite directions due to the lateral inversion in the ancilla preparation setup. This leads to a shift in the
upconversion frequency of $\Delta\omega\tsubs{up} \approx \alpha f\theta_x$ to first order, whilst the zero shear
position is not affected. Due to the low (near normal) diffraction angle from the grating the variation of the shear
slope $\alpha$ is negligible.
We tested these conclusions by recording 50 single shot calibration traces and determining the calibration parameters
shear slope $\alpha$, zero shear row and upconversion frequency $\omega\tsubs{up}$ as shown in
Figure~\ref{fig:calmultishots}. Confirming our findings we find that $\alpha$, and the $x$-origin are constant within
the uncertainty of our fits. The upconversion frequency $\omega\tsubs{up}$ varies by \unit[0.0239]{mrad/fs} consistent
with typical beam pointing fluctuations of our chirped pulse amplification (CPA) system.

\section{Demonstration}
\label{sec:data-accq-reconstr}

In this section we show the characterization of pulses from a home-built CPA system. The
laser delivers \unit[60]{fs} pulses centered around \unit[800]{nm} at a repetition rate of $\unit[2]{kHz}$.
\begin{figure}[htbp]
  \centering
  \includegraphics[width=0.9\textwidth]{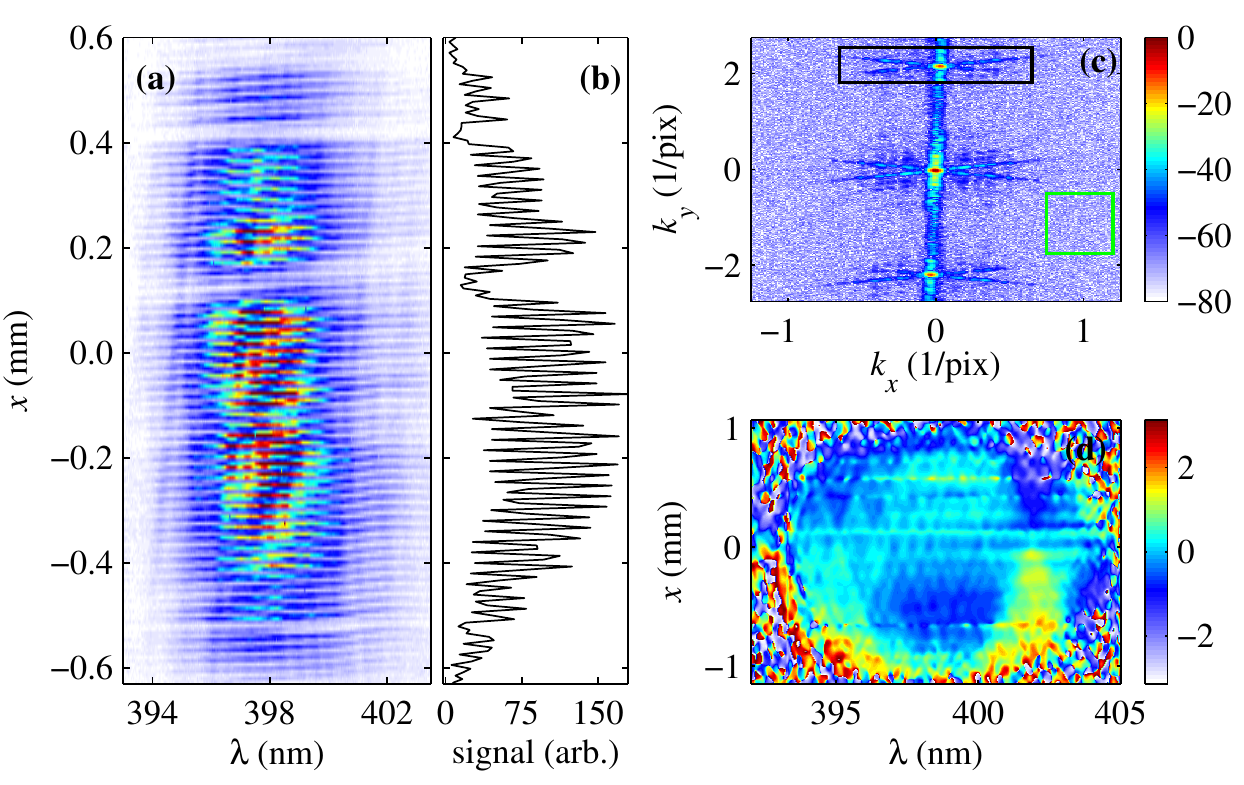}
  \caption{(a) Typical single shot SEA-CAR-SPIDERgram for a \unit[54]{fs} pulse. (b) Spatial lineout at center
    wavelength showing the spatial carrier fringes. (c) 2D-DFT of the interferogram (80 dB scale). The black box
    indicates the sideband filter, and the green box indicates the signal-free region used to determine the noise
    amplitude.  (d) $\Gamma(\omega,x)$ after filtering ($2\pi$ scale). The spatial carrier has been removed for
    clarity.}
 \label{fig:trace}
\end{figure}

A typical single shot SEA-CAR-Spidergram is shown in Fig.~\ref{fig:trace}(a). The regions of low intensity are caused by
dead rows in our detector and imperfections in the upconversion crystal. Confirming the calibration of the upconversion
frequency, the signal is centered around $\lambda=\unit[398]{nm}$ which is equivalent to $\omega_0 + \omega_{\text{up}}
= \unit[(2.345+2.388)]{rad/fs}$ or $\unit[397.98]{nm}$ where $\omega_0$ is the pulse centre frequency.

To aid isolating the interferometric component from the baseband term in the Fourier-filtering procedure we use a high
spatial carrier of $k_x=\unit[2.15]{rad/pix}$ (where $k_x=\pi$ is the Nyquist rate) resulting in fine fringes. A lineout
at $\lambda=\unit[398]{nm}$ is shown in Fig.~\ref{fig:trace}(b). The two-dimensional discrete Fourier transform of these
data is shown in Fig.~\ref{fig:trace}(c); the black rectangle indicates the filter region. The phase of the filtered sideband is shown 
in Fig.~\ref{fig:trace}(d) with the spatial carrier removed for clarity. The symmetry of the signal about $x=0$ is evident.
\begin{figure}[htb]
  \centering
  \includegraphics[]{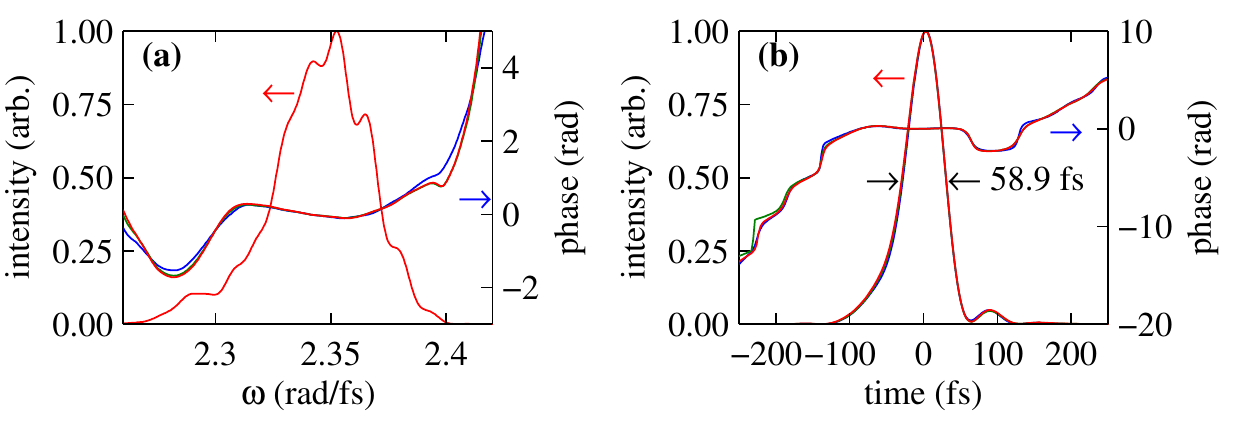}
  \caption{Example reconstructions for different shears $\Omega=12.07, 13.27, \text{and}\,\unit[18.08]{mrad/fs}$. (a) Spectral
    domain with amplitude extacted from the trace at $x=0$ and the 3 phases correspondiong to each shear. (b)
    Temporal domain: temporal intensities and phases for each shear.}
  \label{fig:exampleshears1d}
\end{figure}
After subtracting the calibration phase and extracting the amplitude as described in section~\ref{sec:sea-car-spider} we
obtain the spectral phase for each $\Omega$ present in the data.  The resulting reconstructed spectral phase
$\phi(\omega)$ is displayed in Fig.~\ref{fig:exampleshears1d}~(a) for a subset of the shears present in the
SEA-CAR-Spidergram. The reconstructions show good agreement which we quantify below. Fig.~\ref{fig:exampleshears1d}~(b)
shows the corresponding temporal intensities.

\section{Robustness and precision of the pulse field reconstruction}
\label{sec:robustn-prec-eval}
In this section we investigate the uncertainty of our measurements, showing that SEA-CAR-SPIDER is accurate and
precise. Several features of the SEA-CAR-SPIDER enable a convenient and empirical evaluation of the uncertainty. Its
single-shot nature enables the random errors to be evaluated by studying the statistics of an ensemble of sequential
acquisitions, whilst the presence of systematic errors caused by spatial nonuniformities is revealed by variation
between the reconstructions from different shears. The magnitude of the errors resulting from various misalignments,
instabilities or miscalibrations can also be conveniently calculated due to the simple analytic reconstruction
algorithm.

\begin{figure}[htb] \centering
  \includegraphics[width=0.85\textwidth]{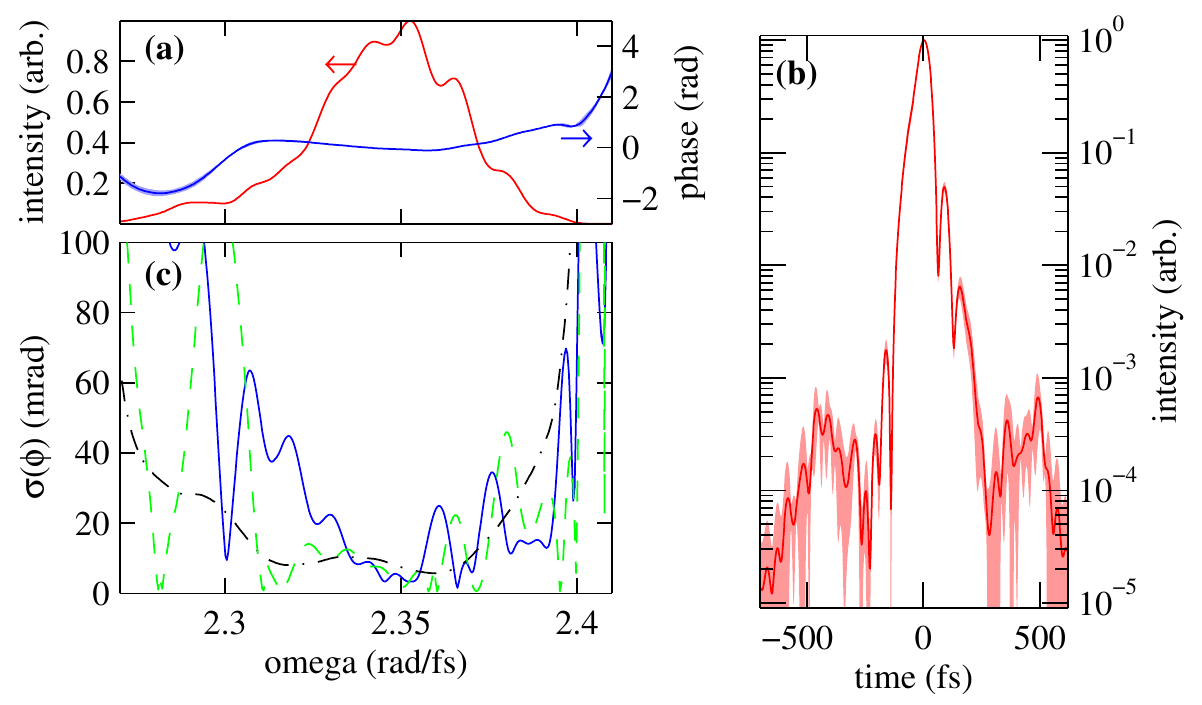}
  \caption{Reconstruction of a 59\,fs pulse for shears from 8.4 to 14.4\,mrad/fs. (a) Spectral domain; intensity (red,
    left axis) and mean phase (dark blue line, right axis) and $\pm$ 1 standard deviation (light blue region) across the
    shears. (b) Temporal intensity; mean (red) and $\pm$ 1 standard deviation interval (light red region) across the shears. (c) RMS
    phase variations magnified: over all shots at a single shear (black dash-dotted), over all 12 shears (blue, solid), and expected variation in the reconstructed phase resulting from fluctuations in the upconversion frequency (green dashed).}
 \label{fig:multipleshears}
\end{figure}

We first evaluated the magnitude of the random errors by examining the statistics over $N=50$ single shot traces.  We
reconstruct each shot independently at $K=12$ different shears $\Omega_k$, ranging from \unit[8.4]{} to
\unit[14.4]{mrad/fs}, yielding a set of reconstructed phases $\phi_{k,n}(\omega)$ from which shot-to-shot statistics can
be obtained. We calculated the mean square shot-to-shot spectral phase variation $\sigma^2_k(\omega) =
\langle[\Delta\phi_{k,n}(\omega)]^2\rangle_n$, where $\Delta$ denotes deviation from the mean value, and angle braces
denote the expectation value over an ensemble which is specified by the subscript. We find that $\sigma_k(\omega)$ is
less than \unit[0.05]{rad} across the full-width at 10\% maximum bandwidth, but did not vary greatly between shears. It
is shown for $\Omega = \unit[10]{mrad/fs}$ by the black dash-dotted line in Fig.~\ref{fig:multipleshears}~(c).  The
shot-to-shot variation of the temporal RMS pulse duration was less than \unit[0.33]{fs} for the different shears with
smaller variations for larger shears. 
We also calculated the RMS field variation $\varepsilon_k$~\cite{Dorrer_2002_JOSAB_Accuracy-criterion-for-ultrash} for
each shear, which was typically around \unit[0.0087]{}.

We attribute this shot-to-shot fluctuations to at least two sources. The phase error resulting from camera shot noise
can be evaluated by the following procedure: the noise density in the Fourier domain is evaluated by examining a region
with no signal, as shown by the green rectangle in Fig.~\ref{fig:trace}(c). The total noise energy in the filtered
signal is then estimated by multiplying this density by the area of the sideband filter. By Parseval's theorem, this is
equal to the noise energy in $D(\omega,x)$ after inverse Fourier transforming. Assuming the noise is uniformly spread
over the $(\omega,x)$ domain, the noise variance $\eta_D^2=\langle[\Delta D(\omega, x)]^2\rangle$ on the samples of
$D(\omega,x)$ is then estimated. Here, we find a peak SNR of $\max D(\omega,x) / \eta_D =536$. By contrast, before the
filtering, the peak SNR was 49. The noise on $\Gamma$ is then
$\eta_{\Gamma(\omega,x)}^2=\mytextfrac{\eta_D^2}{2|D(\omega,x)|^2}$~\cite{Austin_2009_JOSAB_High-precision-self-}
because in regions of low intensity the phase is less well defined. Assuming these errors are independent, the variance
on the reconstructed phase resulting from the concatenation algorithm is the sum of their variances at the sampled
points $\omega_n = \omega_0 + n\Omega$ for $n=0,1,\ldots$. Therefore, the variance of the reconstructed phase difference
between $\omega_m$ and $\omega_n$ is $\gamma^2 =
\langle\{\Delta[\phi(\omega_m)-\phi(\omega_n)]\}^2\rangle=\sum_{k=n}^{m-1}\eta_{\Gamma(\omega_k+\Omega/2,x)}^2$. As an
indicative measure we choose $\omega_n=2.3142$ and $\omega_m=\unit[2.3797]{rad/fs}$ as the lower and upper frequencies
at half the maximum spectral intensity. Here we find $\gamma=\unit[0.02]{rad}$, consistent with the observed
shot-to-shot fluctuations.

Another source of shot-to-shot variation was in $\omega\tsubs{up}$, as described in section~\ref{sec:calibration}. An error
$\Delta \omega\tsubs{up}$ in this quantity leads to a shift in the reconstructed spectrum, giving a phase error of
$\approx \Delta \omega\tsubs{up} \mytextfrac{\partial \phi(\omega)}{\partial \omega}$. Using the RMS value of $\Delta
\omega\tsubs{up}$ obtained from the data in Fig.~\ref{fig:calmultishots}(c), we computed the resulting phase errors
which are shown by the green dashed line in Fig.~\ref{fig:multipleshears}(c). These are consistent with the observed
phase fluctuations (black dash-dotted line).

We now examine the systematic errors. Some of these are revealed by inconsistencies between the reconstructions from
different shears. We computed the mean-square variation over the different shears $\sigma^2(\omega) = \langle [\Delta
\phi_k(\omega) ]^2 \rangle_k$ where $\phi_k(\omega) = \langle \phi_{k,n}(\omega) \rangle_n$ is the shot-averaged phase
reconstructed using shear $\Omega_k$. We have plotted $\sigma(\omega)$ as the blue solid line in
Fig.~\ref{fig:multipleshears}(c). It is of similar magnitude to the shot-to-shot fluctuations. However, in
SEA-CAR-SPIDER, disagreement between the reconstructions from different shears arises from spatial inhomogeneities in
the beam. We verified this by recording a 2D spectrum of the test-pulse and extracting the variation of the spectral
moments accross the beam $x$-axis.  This can also be directly extracted from the data trace by comparing the extracted
amplitudes at different shear positions with the expected amplitudes $|E(\omega+\alpha x_k)||E(\omega-\alpha x_k)|$,
which can be calculated from the spectrum $|E(\omega)|$ extracted at the $x$-origin. Since all temporal characterization
techniques involve either spatial averaging or spatial selection, the ability to quantify the spatial variation is
advantageous.

The final systematic errors we considered arose from the accuracy of our calibration of $\alpha$ and the $x$-origin.
An error $\Delta\alpha$ leads to a fractional shear error of $\mytextfrac{\Delta\Omega}{\Omega} =
\mytextfrac{\Delta\alpha}{\alpha}$ which we measure as \unit[0.3]{\%}, indicated by the error bars in
Fig.~\ref{fig:calmultishots}(a). The resulting phase error is also fractional, scaled by the same amount, and is negligible
for the data presented here. The error in the $x$-origin is more difficult to analyse, as it affects both the shear and the calibration phase.  However, performing the
reconstruction algorithm on this data with a range of different values for the $x$-origin did not affect the reconstructed phase.  We
note that had we used the calibration method using the broadband signals, these errors would be significant.

Since averaging of the statistically independent shot-to-shot fluctuations reduced the random errors to significantly
below the systematic errors, the latter define the ultimate precision of our measurement. For an intuitive visualization
of their magnitude, we plot the temporal profile (shown in Fig.~\ref{fig:multipleshears}~(b) on logarithmic scale)
corresponding to the grand average phase $\langle\phi_k(\omega)\rangle_k$ with the variation between the shears
indicated by the shaded area. We achieve a temporal intensity dynamic range of $10^{-4}$.

We also verified the accuracy of the reconstruction by measuring the spectral phase introduced by a \unit[10]{cm}
long piece of BK7 glass which we inserted into the beam.
We measured a quadratic phase of $\varphi_2=\unit[4400\pm26]{fs^2}$, which differs from the theoretical value 
$\unit[4442]{fs^2}$
 by $\unit[2]{fs^2}$.

\section{Conclusion}
\label{sec:conclusion}
We have demonstrated a novel self-referencing and easily calibrated ultrafast pulse characterization technique. A full
set of data, consisting of the calibration phase, the spectral amplitude of the unknown pulse, and the spectral phase
encoded with multiple shears, is recorded in a single shot. The redundant information contained in each single shot data
trace provides useful material for consistency checks. A high precision shear and upconversion frequency calibration is
straightforward and can be done in one single step for all shears present without the need of any manual or motorized
adjustments. We envisage that this device will be beneficial for its inherent reliability through a lack of moving
parts, its improved precision, and the redundant information in the trace which assists in error determination.

\section*{Acknowledgments}
\addtolength{\parskip}{\baselineskip} This research was supported in part by the European Commission through the
Research Training Network XTRA (contract MRTN-CT-2003-505138) and by EPSRC ER/S24015/OJ. DRA was supported by an Oxford
University Clarendon Fund Scholarship. IAW acknowledges support from the Royal Society.

$^\dagger$Both authors contributed equally to the work presented.

\end{document}